\documentclass[sigconf]{acmart}

\usepackage{booktabs} 

\acmConference[FinTech@KDD'18]{The First Workshop on Data Science In FinTech}{2018}{London, UK}
\acmYear{2018}
\copyrightyear{2018}

\setcopyright{rightsretained}

\begin{document}
\title{Identifying Financial Institutions by Transaction Signatures}

\author{Noa Haas} 
\affiliation{%
  \institution{Intuit}
}
\email{noa_haas@intuit.com}

\author{Yair Horesh}
\affiliation{%
  \institution{Intuit}
}

\author{Shimon Shahar}
\affiliation{%
  \institution{Intuit}
}

\author{Yehezkel S. Resheff}
\affiliation{%
  \institution{Intuit}
}

\begin{abstract}
Financial data aggregators and Personal Financial Management (PFM) services are software products that help individuals manage personal finances by collecting information from multiple accounts at various Financial Institutes (FIs), presenting data in a coherent and concentrated way, and highlighting insights and suggestions. Money transfers consist of two sides and a direction. From the perspective of a financial data aggregator, an incoming transaction consists of a date, an amount, and a description string, but not the explicit identity of the sending FI. In this paper we investigate supervised learning based methods to infer the identity of the sending FI from the description string of a money transfer transaction, using a blend of traditional and RNN based NLP methods. Our approach is based on the observation that the textual description field associated with a transactions is subjected to various types of normalizations and standardizations, resulting in unique patterns that identify the issuer. We compare multiple methods using a large real-word dataset of over $10$ million transactions.
\end{abstract}


\maketitle

\section{Introduction}

Personal financial management (PFM) services and financial aggregators are software applications that collect and bring together information from multiple sources to provide users with a single stop shop for tracking and managing their personal finances \cite{gupta2014personal}. For individuals with multiple bank accounts, credit cards, and utility bills, seeing the big picture and gaining insights into their financial health can be incredibly valuable. Indeed, services of this sort are used by millions of people in the US alone \cite{green2017account}. 

One of the most important types of information collected and analyzed by PFM services are transactions. Bank and credit card transactions are retrieved from financial institutes after users provide the appropriate credentials. These pieces of information essentially sum up to the full financial story pertaining to an individual. However, in order to distill the most relevant insights and suggestions for users, PFMs must fully understand the nature of the observed transactions, their source and meaning. One case of fundamental importance is bank transfers. Across the plethora of Financial Institutes (FIs) in the US, the information consistently retrieved by the service is the date, dollar amount, and a varying length string describing the transaction. These strings are semi human-readable, and in the general case do not include an explicit identification of the issuing FI of the bank transfer \footnote{This information is often available to the customer in the online bank account display, but is not obtained by the PFM due to technical issues.}.

Transaction data includes many types of implicit information that can be extracted using machine learning and data mining methods. Previous research addressed the location of transactions \cite{intuit-pkdd}, and demographic attributes of users \citep{intuit-wsdm}. In this paper we utilize the description string associated with bank transfer transactions (i.e. the \textit{transaction signature}), and treating this as short textual data we apply Natural Language Processing (NLP) and supervised learning techniques to learn the mapping between description strings and the identity of the FI issuing the transfer. 


Description strings are formed by the FIs, presumably, by formatting a template with information regarding a specific transaction. At first glance it might seem that Recurrent Neural Networks (RNNs) and other machine learning approaches are not the right tool for inferring these deterministic mappings (although commonly used successfully for tasks over short strings \cite{limsopatham2016bidirectional,korpusik2016recurrent,yen2018detecting}). The simplicity and accessibility of these methods, and excellent results obtained in our task (see Section \ref{sec:results}) lead us to favor them over more traditional data mining tools designed specifically for finding patterns in strings. 

The rest of the paper is structured as follows: Section \ref{sec:problem} contains a precise problem definition. Next, in Section \ref{sec:methods} the methods are presented, followed by results on a large real-world dataset in Section~\ref{sec:results}.

\begin{table*}
  \caption{Examples of the money transfer transactions and their description by the receiving entities. A small percent of transactions include the sender financial institute explicitly (oversampled here and shown in bold), the rest we attempt to infer from the structure of the string. Numbers and other identifying or private information is censored using Xs. }
  \label{tab:trans-ex}
  \begin{tabular}{cllll}
    \toprule
    date & amount & from  & to & description on receiving end \\
    \midrule
    11.01.16 & 1,000\$    & Bank of America & Bank of America & Online Banking transfer to SAV XXXX Confirmation\# XXXXX \\
    12.01.16 & 1,000\$ & Bank of America & Chase & Online Transfer XXXXX \textbf{fromBofA} main account \#\#\#\#\#\#\#\#XXXX t \\
    01.01.17 & 1,000\$ & Chase & Wells Fargo & \textbf{CHASE} EPAY XXXXX XXXXX <Sender Name> \\
    02.01.17 & 1,000\$ & Wells Fargo & Bank of America & Payment \\
    03.01.17 & 1,000\$ & ING direct & Chase & CAPITAL ONE N.A. CAPITALONE XXXXX WEB ID: XXXXX \\
    \bottomrule
\end{tabular}
\end{table*}

\section{Problem Definition}
\label{sec:problem}
We formalize the problem as the recovery of the identity of the formatter program running by the transaction issuer. Consider a transaction $T$ with a set of attributes $\{a_i(T)\}$. The \textit{issuing formatter} $f_{s}$ running by the sending FI is a mapping from  $A_s \subset \{a_i(T)\}$ to a string. The \textit{receiving formatter} $f_{r}$ running by the receiving FI is a mapping from the received string and $A_r \subset \{a_i(T)\}$ to the final description string we observe. Thus, we observe the string:

\[ f_r(f_s(A_s),A_r) \overset{\Delta}{=} \Pi_{(s,r)} (T) \]

From an in-depth exploration of the data we conclude that the $f_r$ formatters leave much of the structure produced by $f_s(A_s)$ intact. Furthermore, we observe that transactions originating from difference FIs have uniquely identifying patters, albeit this is a many-to-one relation (see Table~\ref{tab:trans-ex}). 

Given many transaction strings $\Pi_{(s_i,r_i)} (T_i)$ our goal is to recover the pairs $(s_i,r_i)$ of the sending and receiving formatters that produced them. Note that since one side of the transaction is known (this is the financial institute in which we saw this transaction), we only need to infer the other side of the transaction. In this paper we concentrate on incoming transactions, where $r$ in known, and infer $s$ from the transaction strings. 

\section{Methods}
\label{sec:methods}

\subsection{Generating the Labeled Dataset}
\label{subs:dataset}

Data used for this work was collected by a large financial data aggregation service. During registration, users provide credentials that allow us to continuously obtain transaction data from over $25,000$ financial institutions including banks and credit card companies. A record describing a transaction typically contains the date of the purchase, a dollar amount, and a \textit{description string} explaining the nature of the transaction. Overall, available data contains over 15 billion transactions per year, arriving from over 10 million users. This represents several percent of all private transactions in the US. In our experiments, we use slices of this data pertaining to money transfer between known financial institutes. All experiments were conducted with data from the year starting November 2016.

In order to generate a labeled dataset of transactions between known financial institutes we use transactions for which both sides are visible to the data aggregation service. More specifically, we concentrate on transactions where both the source and the destination are within the same user account. In such cases we are able to obtain the identity of both financial institutes, as well as the descriptions produced by both of them. 

The labeled dataset obtained this way contains $10.87$ million records, from $88,000$ users. Each record consists of the name of the sending and receiving financial institute, a dollar amount and date, and the description of the transaction as recorded both by the sender and the receiver (see illustrative examples in Table \ref{tab:trans-ex}). Experiments reported here were conducted using a random sample of $500,000$ records from this dataset.

\subsection{Tokenization, Feature Crafting, and Models}
\label{subs:tokens-features}

Description strings were tokenized using a standard (NLTK \cite{bird2004nltk}) tokenizer, limited to a dictionary of size $10,000$. No text pre-processing was preformed, other than replacing digits with Xs (this was done so that tokens representing number lengths would be formed to replace individual numbers). 

In addition to the tokenized representation of the description strings, additional hand-crafted features describing textual patterns that are not expressed as token were computed. These features include indicators (ex. is all the string upper case?), and more complex regular expression patterns found to be useful for this task. 

\noindent We compare the following classification methods and baselines:

\begin{itemize}
\item max-label baseline: as a baseline for all other methods we use the proportion of data from the largest FI in the set under consideration. 

\item logistic-raw: logistic regression on the distribution of tokens only.

\item logistic-features: logistic regression with the additional computed features and the raw token distributions combined.

\item LSTM/GRU: The model structure for all RNN based methods used here consists of a token embedding layer (in all cases the embedding size is $20$), followed by a single LSTM or GRU layer. The final output of the RNN is then fed into a cascade of $2$ dense layers, and a softmax readout of the identity of the financial institute. In the RNN setting only the tokenized sequence is used (with no hand-crafted features). Description string length was limited to $20$ tokens (longer ones were truncated). 
\end{itemize}

\section{Results and Discussion}
\label{sec:results}

\begin{figure*}
\centering{}
\includegraphics[width=\linewidth, trim=0cm 0 0 0, clip]{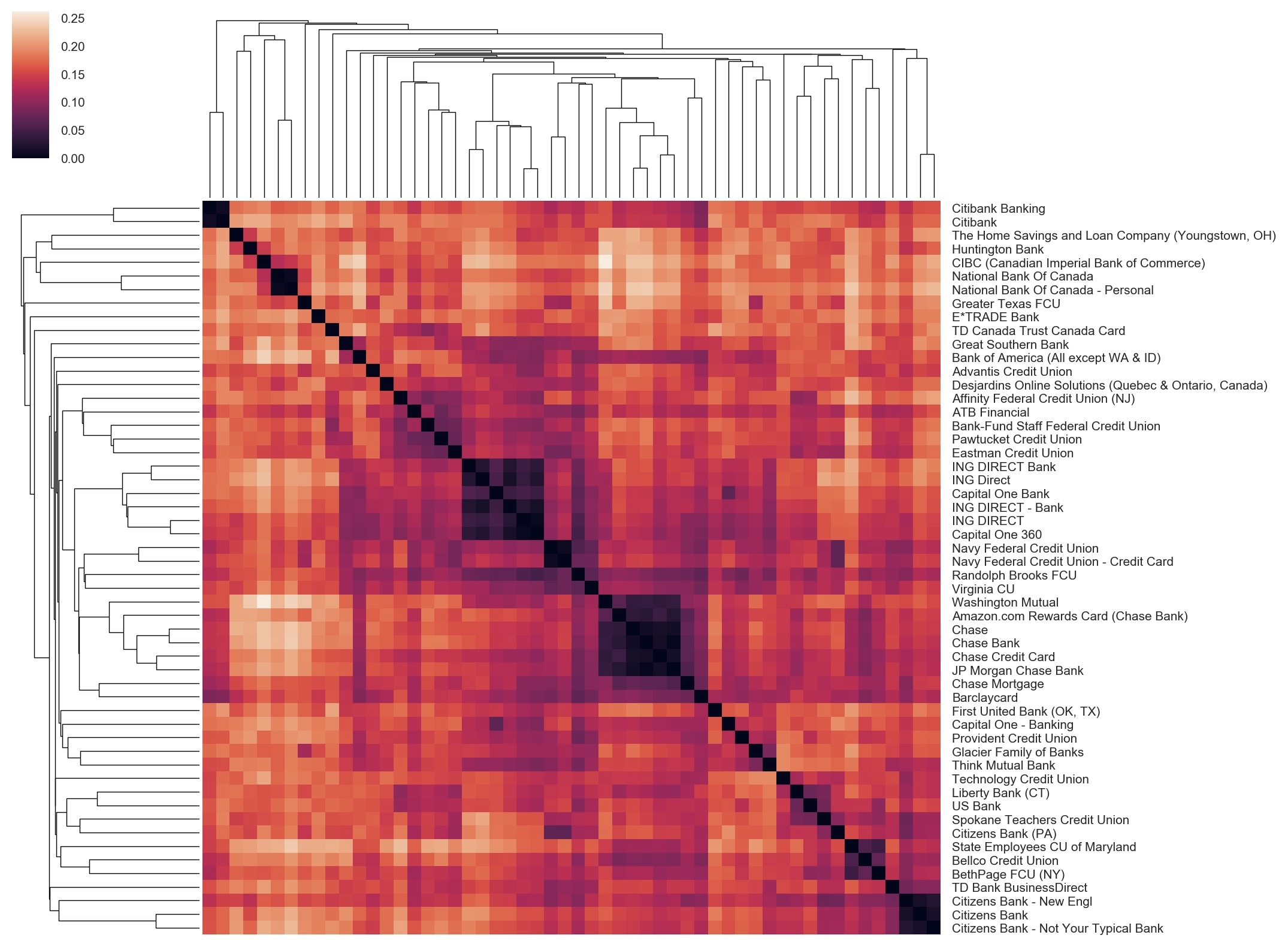}
\caption{A clustering of FIs based on distance between token distribution in description strings.}
\label{fig:bi-clustering}
\end{figure*}

\begin{table}
  \caption{Accuracy of the various methods tested in prediction of financial institute from transaction description. Columns show results when limiting to the most common $10$ / $100$ / $1000$ FIs.}
  \label{tab:results-moethods}
  \begin{tabular}{lccc}
    \toprule
    Method            & $10$-class & $100$-class & $1000$-class   \\
    \midrule
    Baseline-max      & 28.80 & 20.51 & 16.90 \\
    Logistic-raw      & 82.01 & 72.72 & 67.47 \\
    Logistic-features & 82.47 & 73.26 & 68.04 \\
    LSTM              & \textbf{99.13} & \textbf{90.81} & \textbf{84.15} \\
    GRU               & 99.11 & 90.72 & 84.13 \\
    \bottomrule
\end{tabular}
\end{table}

In an exploratory phase, we examined the manner in which the distribution of tokens in descriptions reflects the relations between financial institutes in the US. After tokenization we observed the association between financial institutes as pairwise distances with respect to token distribution. Plotting a clustered heatmap of these distances (see figure \ref{fig:bi-clustering}) reveals that the textual data is useful in revealing associations between different banks. 

For example, the token distribution seems to easily capture the relatedness of different branches or devisions of the same bank, as in the case for Citibank and Chase bank (including Amazon award visa which Chase operates). This view of the data also surfaced mergers and acquisitions in the FI market, such as Capital One's acquisition of ING Direct division. Finally, we learned that the descriptions may also generate geographical attributes, as demonstrated by the moderate similarity between CIBC and National Bank of Canada, which are two distinct institutes. The later observation raises the potential of learning more characteristics of financial institutes through their transaction descriptions. This might also imply a limitation on the learnability of the mapping from description strings to FIs. More precisely, it indicates that we are likely to have to rely on structure and deeper features of these strings, and not just the distributions of tokens. This notion is reinforced by the results presented below. 

We test multiple methods for determining the identity of the financial institute from which a transaction originated based on the description of the transaction (See section \ref{sec:methods} for data and model details). Experiments show overall satisfactory results, with classification accuracy ranging from over $99\%$ when only the top $10$ FIs are considered to approximately $84\%$ for the top $1000$ (Table \ref{tab:results-moethods}). In all cases the LSTM based classifier outperformed all other methods, followed closely by the GRU (It is noteworthy that the logistic regression operated on single token distributions and manually crafted features. Multi-grams were not tested for computational reasons). The vast superiority of both RNN based methods (which operate on the raw token sequences) over the logistic regressions which are not able to take the order of tokens into consideration indicates again that the structure of the description string has an important role in determining the identity of the source FI, and not just the actual tokens used.

\begin{figure}
\centering{}
\includegraphics[width=\columnwidth, trim=0cm 0 0 0, clip]{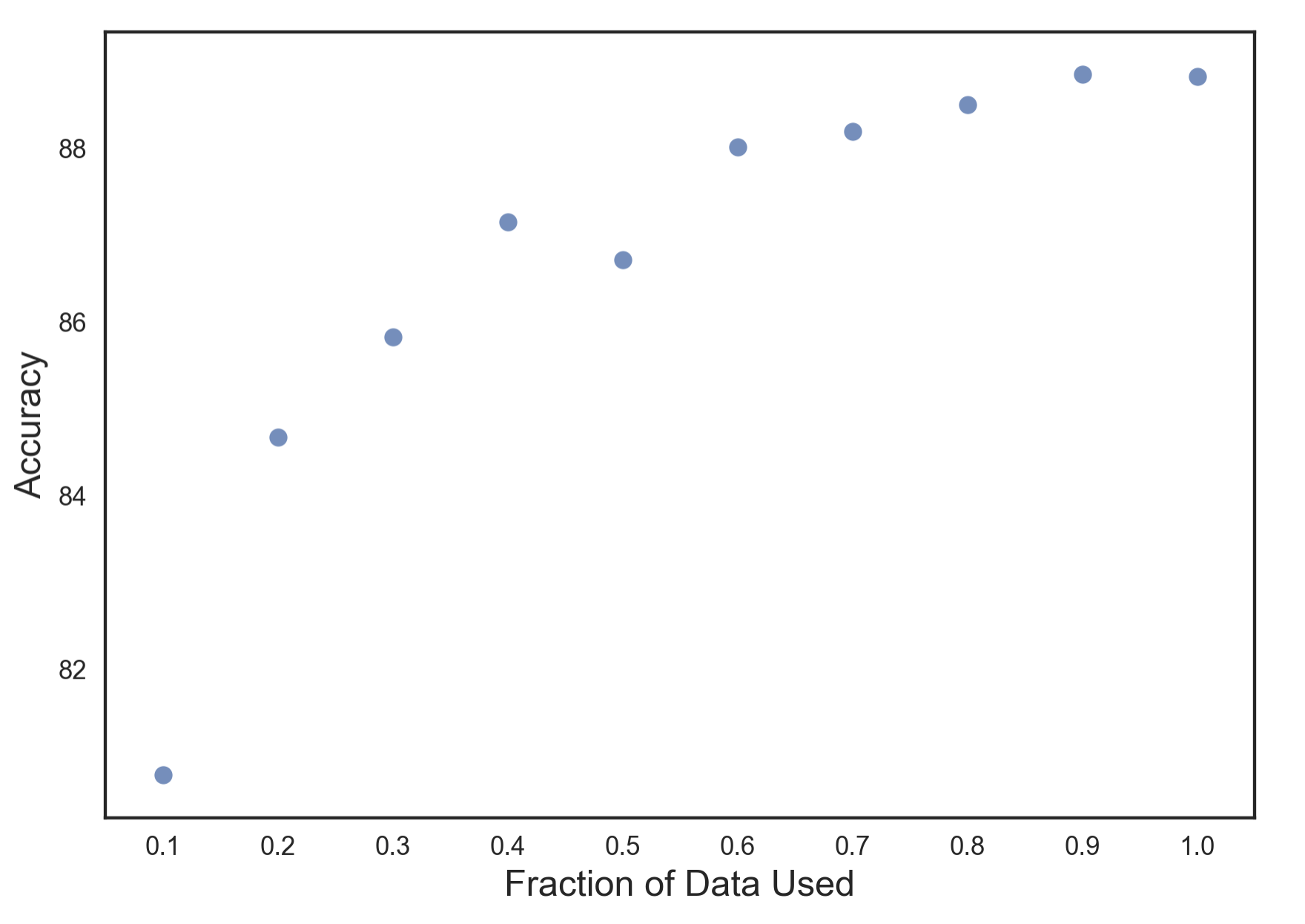}
\caption{Model accuracy as a function of the amount of data used for training. Plot shown for the LSTM based model with a size $50$ representation, and data containing $200$ FI labels. The saturation of the accuracy indicates that the method is unlikely to benefit from additional data. }
\label{fig:fracdata}
\end{figure}

Since the experiments presented in this paper are conducted on a subset of the available data, we test to determine the sensitivity of the classification results to the amount of training data used. Results for the LSTM based method in the $200$ FI setting (Figure \ref{fig:fracdata}) show that performance reaches a plateau at $60\% - 100\%$ of the data used in practice, indicating that the use of additional data would be unlikely to achieve better results. We do not however rule out the possibility of utilizing the full amount of data available with more complex models, or when classifying a larger number of FIs, and leave this to future work. 

\begin{figure}
\centering{}
\includegraphics[width=\columnwidth, trim=0cm 0 0 0, clip]{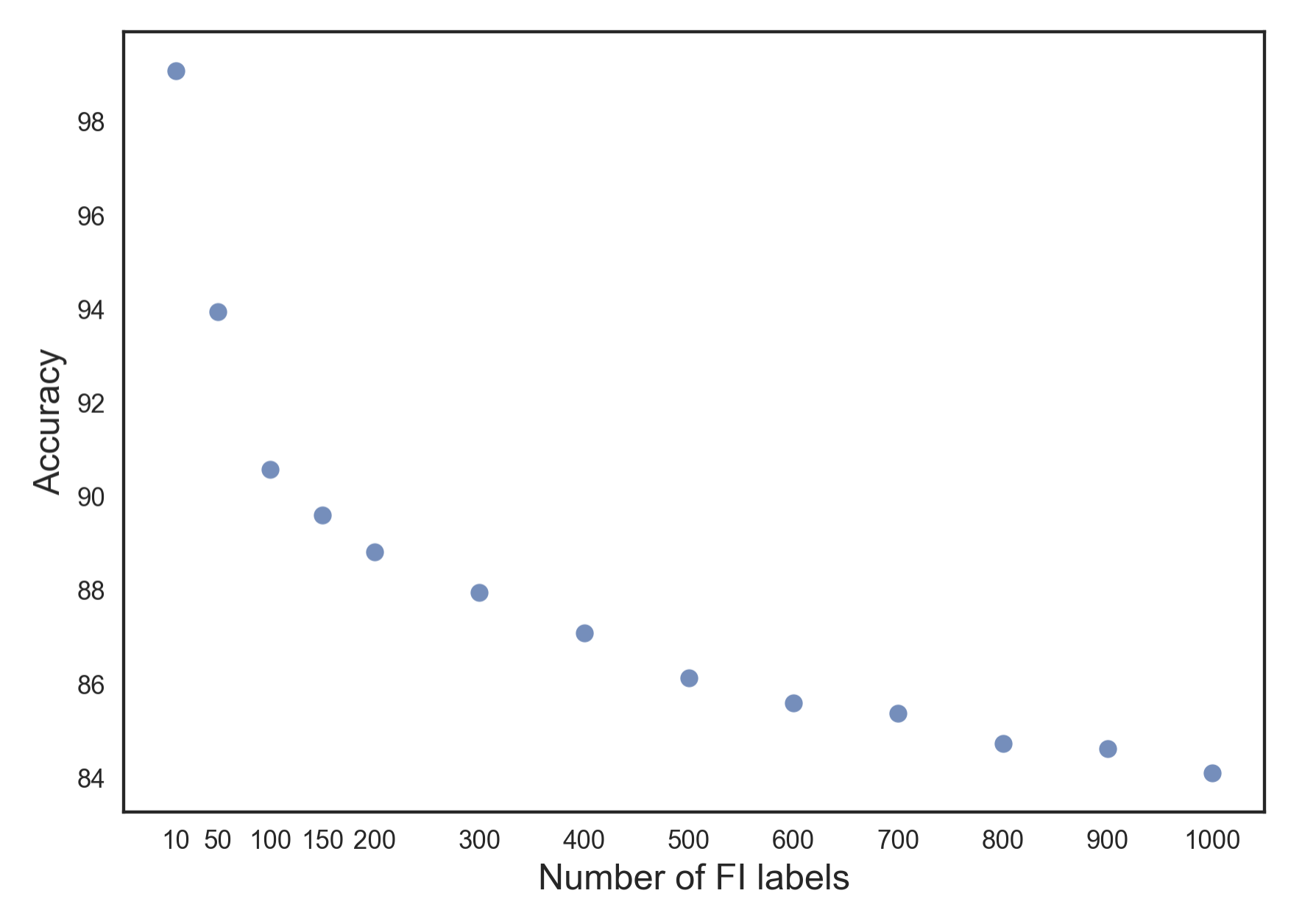}
\caption{Model accuracy as as function of the number of financial institutes (labels) to be classified. Plot shown for the LSTM based model with a size $50$ representation.}
\label{fig:nfi}
\end{figure}

Next we test the trade-off between the number of FIs we classify and classification performance. The US banking system is comprised of tens of thousands of institutions with a long tail distribution of number of customers. In the data used for these experiments the top $10$ institutions are responsible for approximately $50\%$ of all transactions, and the top $1000$ for approximately $95\%$. The decline in performance in the LSTM based method as additional FIs are added follows this structure closely (Figure \ref{fig:nfi}), with a reduction from $99.13\%$ with $10$ FIs to $90.81\%$ for $100$. The decline then slows down, and reaches $84.15\%$ with $1000$ FIs.

\section{Conclusion}
Understanding the source and meaning of transactions is a key component in the ability of financial data aggregators and personal financial management systems to deliver value through deep insights and suggestions. Money transfers are an especially important type of transaction, but the identity of the sending financial institute is not readily available in PFM aggregators systems. 

In this paper we investigate the problem of supervised learning of the identity of a sending financial institute from the description string provided by the receiver. Using word embeddings, RNNs and other methods borrowed from NLP we are able to achieve excellent accuracy on this task, possibly limited only by the multiplicity of banking brands within the same family of banks. Interestingly, RNN methods with the ability to process the order of tokens in the transaction strings vastly outperform linear methods (even when additional hand-crafted features were added to the latter). This finding further supports our original hypothesis that the structure of these strings is tied to issuing FIs, and not merely the distribution of tokens. 

Future work will attempt to enrich the information regarding incoming money transfers beyond the identity of the sending FI by utilizing and extending the methods presented in the current work to recover the structure of description strings and extract the attributes of the transaction embedded within them.

\bibliographystyle{ACM-Reference-Format}
\bibliography{lib}

\end{document}